\documentclass[letter]{aa}
\usepackage{graphicx,txfonts}
\usepackage{hyperref}
\usepackage{subfigure}
\usepackage[section]{placeins}
\usepackage{soul}
\hypersetup{
  colorlinks=true,   
  urlcolor=blue,     
  linkcolor=blue,     
  citecolor=blue
}

\newcommand{\teff}{$T_{\rm eff}$}
\newcommand{\logg}{$\log{g}$}
\newcommand{\vsini}{$\nu\sin{i}$}
\newcommand{\vmic}{$\nu_{\rm mic}$}
\newcommand{\vr}{$\nu_{\rm r}$}
\newcommand{\fph}{$f_{\rm ph}$}
\newcommand{\xhobs}{$\mathrm{\left(\frac{X}{H}\right)_{star}}$}
\newcommand{\xhdisc}{$\mathrm{\left(\frac{X}{H}\right)_{disc}}$}
\newcommand{\xhref}{$\mathrm{\left(\frac{X}{H}\right)_{ref}}$}

\newcommand{\loggf}{$\log{(g_{l}f_{lu})}$}
\newcommand{\lboo}{$\lambda\,$B\"{o}o}

\begin{document}
\title{The refractory fraction of phosphorus in planet-forming discs} 

\titlerunning{Phosphorus}

\authorrunning{S. P. D. Borthakur et al.}

\author{Sandipan P. D. Borthakur\inst{\ref{inst1},\ref{inst2},\ref{inst3}}  \and
        Mihkel Kama\inst{\ref{inst4},\ref{inst3}} \and
        Luca Fossati\inst{\ref{inst1}} \and
        Colin P. Folsom\inst{\ref{inst3}} \and
        Anna Aret\inst{\ref{inst3}} \and \\
        Christiane Helling\inst{\ref{inst1}, \ref{inst2}}
        }

\institute{
        Space Research Institute, Austrian Academy of Sciences, Schmiedlstrasse 6, 8042, Graz, Austria\\
        \email{sandipan.borthakur@ut.ee}\label{inst1}
        \and
        Institute for Theoretical and Computation Physics, Graz University of Technology, Petersgasse 16, 8010 Graz, Austria\label{inst2}
        \and
        Tartu Observatory, University of Tartu, Observatooriumi 1, T\~{o}ravere, 61602, Estonia \label{inst3}
        \and 
        Department of Physics and Astronomy, University College London, Gower Street, London, WC1E 6BT, UK\label{inst4}
    }

\date{Received date 17 June 2026 / Accepted date 28 August 2026}

\abstract 
{Phosphorus (P) is an essential element for life on Earth and a potential tracer of planet formation history. However, there has been no detection of P-bearing molecules in protoplanetary discs so far. Herbig Ae/Be stars constantly accrete matter from their protoplanetary disc, which alters the composition of the stellar photosphere due to their shallow convective, or fully radiative, envelope. The altered surface composition reflects the composition of the accreting matter, and thus that of the inner protoplanetary disc. This accretion contamination of stellar photosphere can persist after accretion has ended in young A and B-type stars (age\,$\lesssim$\,50\,Myr).}
{We aim to quantify the fraction of P locked in dust (the refractory fraction of P) compared to gas in the inner protoplanetary disc around Herbig Ae/Be stars.}
{We measure the stellar parameters and abundance of 5 Herbig Ae/Be stars using optical and UV spectra, to compare their P and Fe abundances. We also used a $\sim\,20\,$Myrs old main sequence B-type star, which has P and Fe abundance estimated from optical spectrum.
Fe is assumed to be completely locked in refractory reservoirs in the inner disc. A parameterised relationship between the stellar P and Fe abundance gives the fraction of P locked in refractory reservoirs.}
{We find the refractory fraction of P in the inner protoplanetary disc to be $\geq96\,\%$ within 95th percentile of the posterior distribution.}
{Consistent with a previous finding in the HD\,100546 system, we conclude most of the P in the inner protoplanetary disc is locked in dust, likely in refractory minerals like schreibersite or apatite. Our result, combined with the low cosmic abundance of P, is also consistent with the lack of infrared and mm-wavelength observations of P-bearing molecules in protoplanetary discs to date.}


\maketitle

\section{Introduction}\label{sec:intro}
Phosphorus (P) is one of the essential elements for life on Earth \citep{1987Westheimer,2015Patel}, but it is also the least studied. The enrichment of P in Jupiter's atmosphere is linked to its solid accretion \citep{2019Oberg}. P-bearing molecules, such as PN and PO, have been detected in star-forming regions \citep{1987Turner,2025Scibelli}. $^{31}$P is the only stable isotope of phosphorus, which is synthesised in massive stars for example via $ ^{16}\mathrm{O} (^{16}\mathrm{O},p)^{31}\mathrm{P}$ and $^{30}\mathrm{Si}(p,\gamma)^{31}\mathrm{P}$ reactions \citep{1995Woosley, 2006Kobayashi, 2013Chieffi}, and released into the interstellar medium (ISM) during supernovae explosions \citep{2013Koo}. During star formation, stars and their surrounding discs inherit this P from the ISM, enabling it to be delivered to planets.
P-bearing minerals like schreibersite ((Fe, Ni)$_3$P) and apatite (Ca$_5$(PO$_4$)$_3$A where A is F, Cl or OH) have also been detected in the asteroid and cometary material in our solar System. On Earth, most P is locked in apatite \citep{2023Krijt}, but the early Earth could have had P delivered from meteorites in the form of schreibersite \citep{2021Walton}. 

There has been no detection of P-bearing molecules in protoplanetary discs so far. Upper limits on the gas-phase P abundance in the protoplanetary disc of Herbig Be star, HD\,100546, combined with the P abundance in the stellar photosphere, suggested P to be completely locked in dust \citep{2025Kama}. The P abundance in the stellar photosphere is highly depleted compared to solar abundance, but the amount of depletion is similar to that of refractory elements like Iron (Fe; \citealt{2025Kama}). Instead, the photospheric abundance of volatile elements such as carbon and oxygen is solar \citep{2016Kama,2025Kama}. Stars with lower abundance of refractory elements and solar-like abundances of volatile elements are also called \lboo\ stars.

\begin{figure}
\includegraphics[width=0.95\linewidth,trim = 1cm 0cm 1cm 0cm]{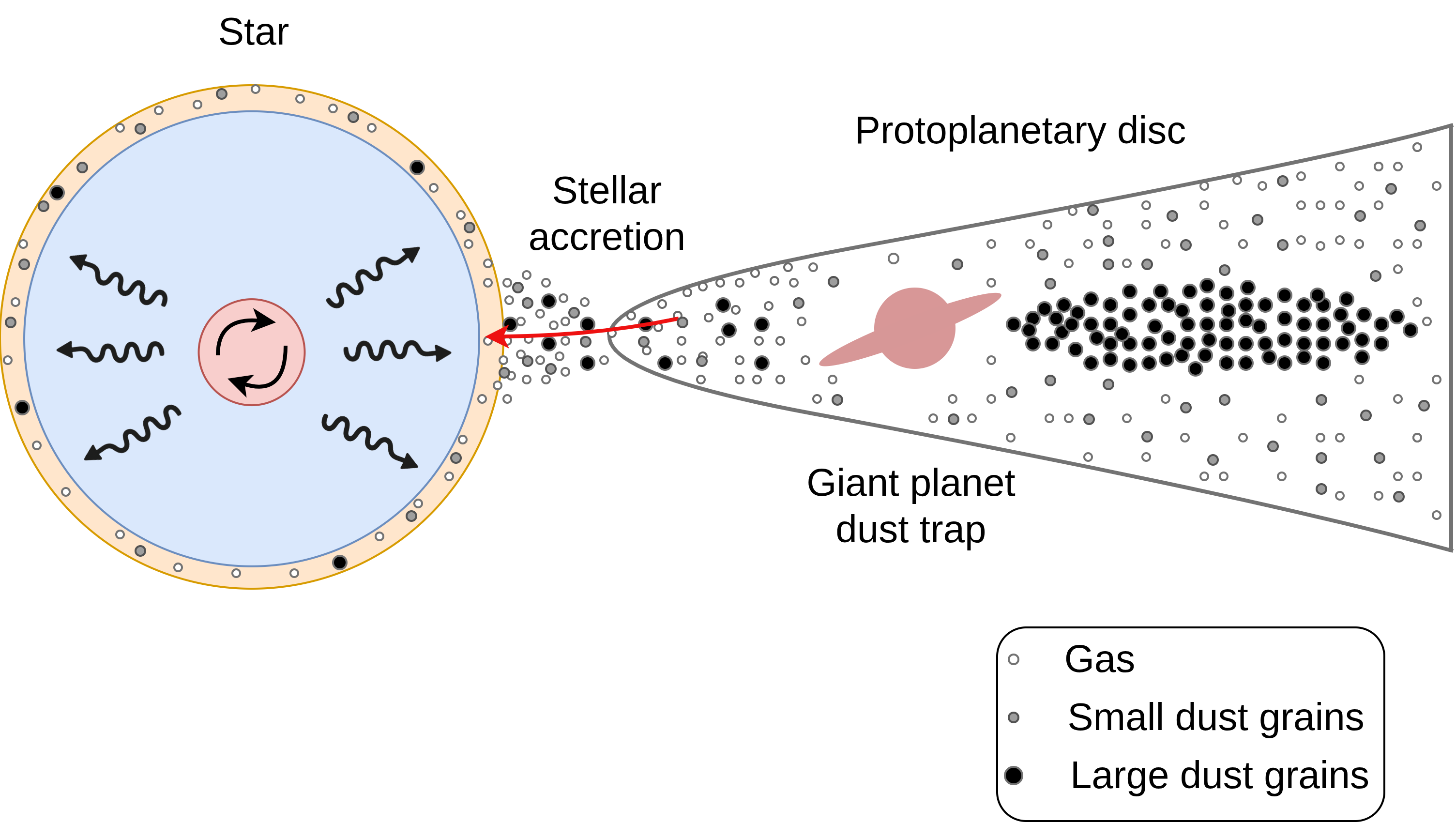}
 \caption{Schematics of a HAeBe star accreting material from its protoplanetary disc with an ongoing giant planet formation. The giant planet opens a disc gap, preventing the larger dust grains from moving inwards. The HAeBe star has a convective core (red), a radiative envelope (blue), and the stellar photosphere (yellow). }
 \label{fig:schematics}
\end{figure}

Herbig Ae/Be stars (HAeBe stars; stellar mass, $M_*\,\gtrsim\,1.4M_{\odot}$) constantly accrete material from their protoplanetary discs. These stars either have a shallow convective envelope or a fully radiative envelope, in which rotational mixing dominates \citep{Jermyn2018StellarPlanets,2020Hoppe}. Whereas cooler stars have deep convective envelopes \citep{2020Hoppe}. Any accretion from the protoplanetary disc thus alters the composition of the HAeBe stellar photosphere due to slower photospheric mixing \citep{Jermyn2018StellarPlanets}. The star accretes gas, as well as small ($\mu$m-sized) and large (mm- to cm-sized) dust grains. The distinction between small and large is made by the strength of coupling, with small grains behaving as large gas particles. If a gap opens in the disc due to giant planet formation, the larger dust grains get trapped at the outer edge of the gap due to gas pressure maxima, whereas the inward gas flow remains unaffected \citep{2012Pinilla, 2012Zhu}. The star accretes a lower amount of larger dust grains if there is a disc gap, and thus also a lower amount of any element mostly locked in dust. Figure\,\ref{fig:schematics} shows a schematic diagram of the above-described process. The P depletion in the stellar photosphere of HD\,100546 indicates that the P in the inner disc is mostly locked in dust, and the inner disc of HD\,100546 is highly depleted in dust \citep{2025Kama}. 

Young A and B-type Stars (YABS; age $\lesssim$ 50 Myrs) are stars which have recently transitioned from their protoplanetary disc hosting phase to either a debris disc phase or no disc phase. The accretion rate from debris discs ($\sim10^{-11}M_{\odot}$/yr; \citealt{2017Kral, 2025Borthakur}) is too low to have a detectable accretion contamination on their stellar photospheres, but YABS can still hold the earlier accretion signature from their protoplanetary disc accretion \citep{2025Borthakur}.

In this paper, we aim to use the depletion of P in stellar photospheres, due to variable amounts of dust accretion onto HAeBe stars, to quantify the fraction of P locked in refractory reservoirs in the inner protoplanetary discs. We will analyse a sample of HAeBe stars (including one YABS from \citealt{2026Olbermann}) to estimate the refractory fraction of P. This idea has been previously used to quantify the inner disc refractory fraction of sulfur to be $89 \pm 8\%$ \citep{2019Kama} by comparing the S abundance of a sample of HAeBe stars with their Fe abundance, where it is assumed that Fe is completely locked in refractory reservoirs. A similar analysis for oxygen reveals a refractory fraction of $2 \pm 2\%$, which means that, within the uncertainties, all O in the inner disc can be accounted for by easily evaporated volatiles \citep{2019Kama}. 

\section{Sample selection, data and analysis}\label{sec:sample}
The targets, the instrument used to obtain the data, and the spectral resolution of the instrument are listed in Table\,\ref{tab:sample_info_table}. The strongest, and only reliably detectable P lines in early F- to late B-type stars lie between $\sim1400 - $2200\,$\AA$ \citep{2011Landstreet, 2025Kama}. We used \texttt{astroquery} to identify objects in the Mikulski Archive for Space Telescopes (MAST) archive\,\footnote{\texttt{\href{https://mast.stsci.edu/search/ui/\#/hst}{https://mast.stsci.edu/search/ui/\#/hst}}} with Hubble Space Telescope (HST) Space Telescope Imaging Spectrograph (STIS) or Cosmic Origins Spectrograph (COS) observations. We excluded objects classified as a supernova, white dwarf, Wolf-Rayet star, cataclysmic variable, or extragalactic object in the \texttt{OTYPE} parameter and only selected objects with spectral types between F0 and B0 using the \texttt{SPTYPE} parameter. We selected stars identified in the literature as HAeBe stars \citep{Acke2004ChemicalStars,2012Folsom, 2015Fairlamb,2021GuzmanDiaz, 2015Kama, 2019Kama} whose UV spectra covered $\sim1400 - 2000\,\AA$, for which archival optical spectra were available, and whose STIS or COS observations were of medium or high spectral resolution. We excluded stars with no satisfactory UV spectral fit to estimate the P abundance. Based on the above criteria, we finalise a sample of five HAeBe stars which we used for our analysis. We also included one YABS HD\,37356 with an age of  $20.2\pm 0.6$ Myrs, which has detailed abundance estimates from an optical spectrum, including non-LTE-corrected P-abundances \citep{2026Olbermann}. HD\,37356 show \lboo\ chemical peculiarity.

We first estimated the stellar parameters and abundances of our sample from archival optical spectra. The spectra are taken from one of the two following spectrographs: Echelle SpectroPolarimetric Device for the Observation of Stars (ESPADONS; \citealt{2014Petit, 1997Donati}) at the Canada-France-Hawaii Telescope, or Fiber-fed Extended Range Optical Spectrograph (FEROS) at the MPG/ESO 2.2-metre telescope. Table\,\ref{tab:sample_info_table} lists the optical spectrograph used for each star and its spectral resolution. The stellar parameters and abundances are listed in Table\,\ref{tab:sample_parameter_table} and \ref{tab:sample_abundance_table} and the details about the analysis are described in Appendix\,\ref{appendix:optical analysis}. The optical spectral fits for the stars analysed here are shown in Figure\,\ref{fig:optical_PPD_fit}. We visually selected wavelength regions in the UV spectra to estimate P abundances, avoiding regions where additional circumstellar absorption or emission lines were observed. The exact P-lines used and estimated abundance from that line for all the sample stars are listed in Table\,\ref{tab:sample_P_abundance_table}. The details about the P abundance analysis are described in Appendix\,\ref{appendix:uv analysis}.

\subsection{Additional sources of uncertainty on the P abundance}
To the best of our knowledge, no non-local thermodynamical equilibrium (NLTE) corrections for P lines in the UV are available in the literature for early-type stars. \cite{2026Andrievsky} has provided NLTE corrections for two P~\textsc{i} lines at 2135.46 and 2136.18\,$\AA$ uptill 6750\,K. Extrapolating the grid to the temperature range of our stars suggests NLTE corrections of $\sim$0.1–0.2 dex, comparable to the uncertainties in our derived abundances. Although the NLTE corrections for the P lines used in our analysis do not necessarily have the same values, they may be of similar magnitude.

The accretion luminosity in HAeBe stars adds an additional continuum spectrum, also called veiling, on top of the stellar continuum. The veiling effect in the optical band is negligible \citep{2012Folsom}, but may be more relevant in the UV. After fitting the continuum and P abundance to the UV spectra, we visually verified that the synthetic spectra reproduced the observed spectra around the P lines. Even if the veiling effect is significant, it can be locally compensated for during the continuum fit, given the larger uncertainties in our P abundances.

\begin{figure}
\resizebox{\hsize}{!}{\includegraphics[trim=0.4cm 0.3cm 0.0cm 0.0cm]{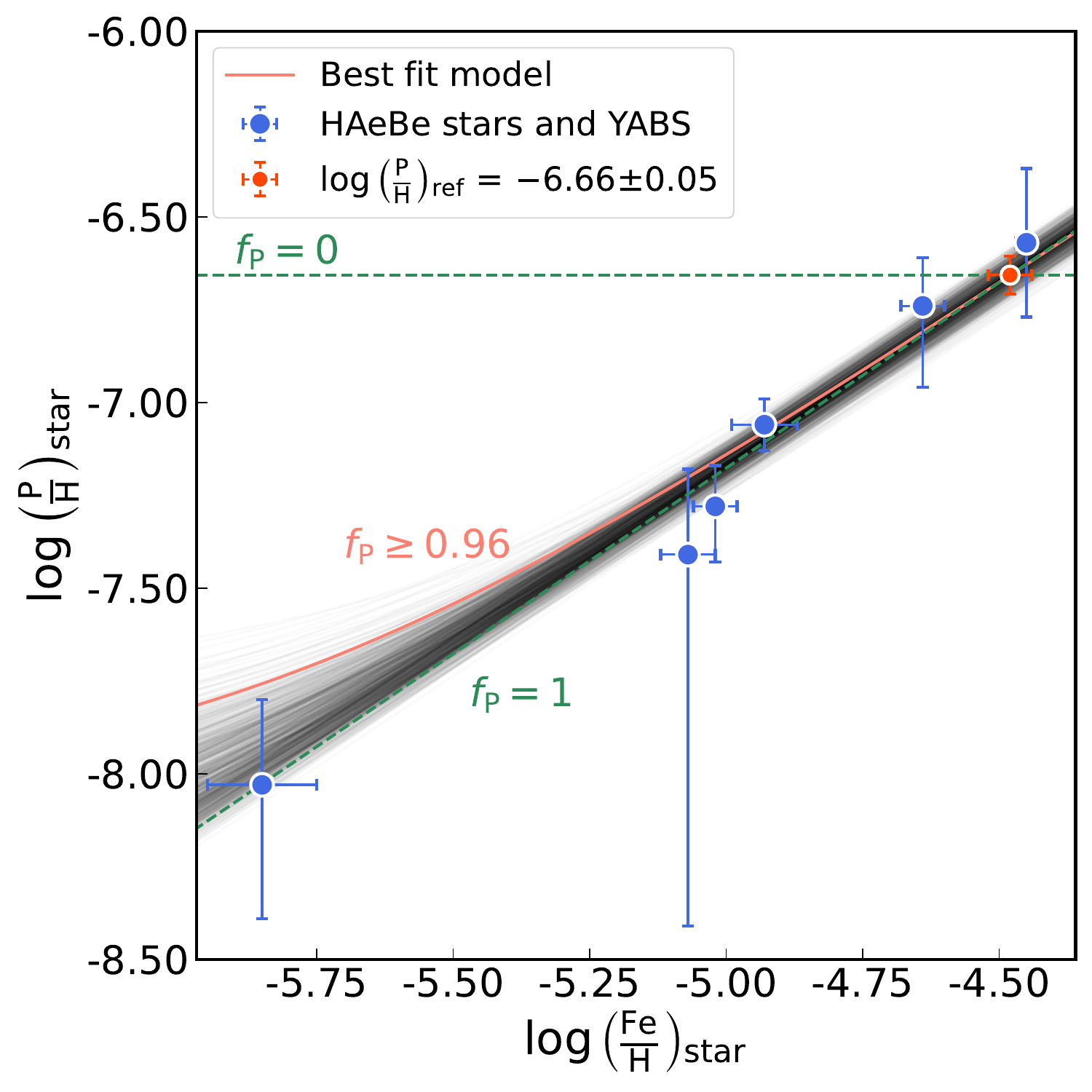}}
 \caption{P abundance as a function of Fe abundance, for the Herbig Ae/Be stars and YABS in the sample, shown in $\log$ abundance. The grey curves represent the posterior distribution of the linear fit to linear abundances (Eq.\,\ref{eqn:linear_PFe_star_relation} \& \ref{eqn:slope_and_intercept_PFe}). The pink curve shows the lower limit of the posterior distribution at the 95th percentile. The horizontal and diagonal green dashed lines represent the expected line if $f_{\rm P} = 0$ and $f_{\rm P} = 1$, respectively. The red datapoint represents the reference Fe and P abundance.}
 \label{fig:P_vs_Fe plot}
\end{figure}

\subsection{Calculation of refractory fraction of phosphorus}
To quantify the fraction of P locked in refractory reservoirs in the inner protoplanetary disc, we followed and extended the methodology described by \citet{2019Kama}. All the abundances of the form $\rm \left(\frac{X}{H}\right)$ are in absolute number density ratios and not in logarithmic scale unless otherwise mentioned. The measured elemental photospheric abundance, \xhobs, of a HAeBe star can be represented as
\begin{equation}
    \mathrm{\left(\frac{X}{H}\right)_{star}} = f_{\rm ph}\,\mathrm{\left(\frac{X}{H}\right)_{disc}}+(1-f_{\rm ph})\,\mathrm{\left(\frac{X}{H}\right)_{ref}}\,,
    \label{eqn:contamination equation}
\end{equation}
where $f_{\rm ph}$ is the fraction of inner disc material currently in the stellar photosphere. This fraction depends on the accretion rate and the mixing mechanism at the stellar photosphere. The term \xhdisc\ is the abundance of the element X in the accreting material contaminating the stellar photosphere. This abundance reflects the abundance of the inner disc. The term \xhref\ represents the abundance of the stellar photosphere without any accretion contamination, and typically reflects the bulk composition of the star. 

Since both the star and disc form from the same material, they inherit the same bulk composition. In the disc, some fraction of any given element is in a dust phase, and some fraction is in a gas phase. Processes such as giant-planet formation can deplete dust in the inner disc. The abundance of an element, X, in the accreting material onto the star can then be represented by  
\begin{equation}
    \mathrm{\left(\frac{X}{H}\right)_{disc}} = (1 - f_{\rm X})\,\mathrm{\left(\frac{X}{H}\right)_{ref}} + f_{\rm X}\,\delta_{\rm d}\,\mathrm{\left(\frac{X}{H}\right)_{ref}}\,,
    \label{eqn:x_disc_frac}
\end{equation}
where $f_{\rm X}$ is the fraction of element X in refractories and $\delta_{\rm d}$ is the fraction of dust in the accreting material. The first term represents the gas-phase elemental abundance, and the second term represents the dust-phase abundance in the inner disc.

We can write the abundance of any other element Y in the accreting material with respect to the elemental abundance of X by substituting $\delta_{\rm d}$ from Eq. \ref{eqn:x_disc_frac} as
\begin{equation}
    \mathrm{\left(\frac{Y}{H}\right)_{disc}} = \mathrm{\left(\frac{Y}{H}\right)_{ref}} \times \left[1+\frac{f_{\rm Y}}{f_{\rm X}}\left(\frac{ \mathrm{\left(\frac{X}{H}\right)_{disc}}}{ \mathrm{\left(\frac{X}{H}\right)_{ref}}} - 1\right)\right]\,.
    \label{eqn:x_vs_y_eq1}
\end{equation}

Since all HAeBe stars accrete material from their discs, their \fph\ value is always non-zero. Even for YABS, their \fph\ value will still be non-zero as their previous accretion from their protoplanetary disc still exists in their stellar photospheres \citep{2025Borthakur}. Then we can rewrite Eq. \ref{eqn:contamination equation} as
\begin{equation}
    \mathrm{\left(\frac{X}{H}\right)_{disc}} = \frac{1}{f_{\rm ph}}\mathrm{\left(\frac{X}{H}\right)_{star}} - \frac{(1 - f_{\rm ph})}{f_{\rm ph}}\mathrm{\left(\frac{X}{H}\right)_{ref}}.
    \label{eqn:contamination equation rearranged}
\end{equation}
Replacing the disc abundance for elements X and Y in Eq.\,\ref{eqn:x_vs_y_eq1} by Eq.\,\ref{eqn:contamination equation rearranged}, we get
\begin{equation}
    \mathrm{\left(\frac{Y}{H}\right)_{star}} = m\,\mathrm{\left(\frac{X}{H}\right)_{star}} + c\,,
    \label{eqn:linear_xy_star_relation}
\end{equation}
where
\begin{equation}
    m = \frac{f_{\rm Y}}{f_{\rm X}}\frac{\mathrm{\left(\frac{Y}{H}\right)_{ref}}}{\mathrm{\left(\frac{X}{H}\right)_{ref}}},\ c = \left(1 - \frac{f_{\rm Y}}{f_{\rm X}}\right)\mathrm{\left(\frac{Y}{H}\right)_{ref}}\,.
    \label{eqn:slope_and_intercept}
\end{equation}
Eq.\,\ref{eqn:linear_xy_star_relation} is also satisfied when the stellar abundances of X and Y are replaced by the reference abundance. If we take an element X that is locked completely in refractory reservoirs in the inner disc, then $f_{\rm X} = 1$. 

In this paper, we compare the P abundance with the Fe abundance to estimate the refractory fraction of P, $f_{\rm P}$. We also assume that Fe is completely locked in refractory reservoirs in the inner disc, and thus $f_{\rm Fe} = 1$. Rewriting Eq.\,\ref{eqn:linear_xy_star_relation} and Eq.\,\ref{eqn:slope_and_intercept} we get
\begin{equation}
    \left(\rm\frac{P}{H}\right)_{\rm star} = m\left(\rm \frac{Fe}{H}\right)_{\rm star} + c\,,
    \label{eqn:linear_PFe_star_relation}
\end{equation}
where 
\begin{equation}
    m = f_{\rm P}\frac{\left(\rm \frac{P}{H}\right)_{\rm ref}}{\left(\rm \frac{Fe}{H}\right)_{\rm ref}},\ c = \left(1 - f_{\rm P}\right)\left(\rm \frac{P}{H}\right)_{\rm ref}\,.
    \label{eqn:slope_and_intercept_PFe}
\end{equation}
The calculation of $f_{\rm P}$ from Eq.\,\ref{eqn:linear_PFe_star_relation} is independent of the stellar accretion rate or the exact mixing mechanism. This allows us to also include the YABS sample and combine it with the HAeBe star sample. We measured the $f_{\rm P}$ and $\left(\rm \frac{P}{H}\right)_{\rm ref}$ from an MCMC linear fit in the P vs Fe abundance plane for our sample of stars (shown in Figure\,\ref{fig:P_vs_Fe plot} in logarithmic scale). We set uniform priors for \( f_{\mathrm{P}} \in [0,1] \) and \( \left(\rm \frac{P}{H}\right)_{\mathrm{ref}} \in [-8,-6] \).

\section{Results}\label{sec:results}
The present-day elemental abundances are consistent across the solar neighbourhood within $\sim$ 0.04\,dex \citep{2012Nieva}. We assumed the reference abundances, $\rm \left(\frac{P}{H}\right)_{ref}$ and $\rm \left(\frac{Fe}{H}\right)_{ref}$ are the same for all stars in our sample. The Fe abundance in the solar neighbourhood in logarithmic scale is $-4.48\pm0.03$ \citep{2012Nieva}. We assumed that $\log\left(\rm\frac{Fe}{H} \right)_{\rm ref} = -4.48$. We estimated the two unknowns $f_{\rm P}$ and $\left(\rm\frac{P}{H}\right)_{\rm ref}$ by solving the two equations for $m$ and $c$. Using the P and Fe photospheric abundance of our sample of HAeBe stars and YABS, we get $f_{\rm P} \geq 96\%$ within 95th percentile of the posterior distribution and  $\log\left(\rm\frac{P}{H}\right)_{\rm ref} = -6.66 \pm 0.05$ where the uncertainties are the 5th and 95th percentiles of the posterior distribution. Our estimated reference P abundance, $\log\left(\rm\frac{P}{H}\right)_{\rm ref}$, is similar to the reference P abundance of $-6.64 \pm 0.14$ measured in the solar neighbourhood \citep{2025Aschenbrenner} and also the latest solar P abundance of $-6.65 \pm 0.04$ \citep{2026Amarsi}. The posterior distribution of the fit is shown in Figure\,\ref{fig:corner_plot_pfrac_pref}.

\section{Discussion and Conclusion}\label{sec:discussion_and_conclusion}
We estimated the fraction of phosphorus locked in refractory reservoirs in the inner regions of protoplanetary discs to be $\geq\,96\%$ with a posterior probability of 95\%, respectively. This result shows that almost all the phosphorus in protoplanetary discs is locked in dust or volatile ices, which also explains the lack of any detections of P-bearing molecules in these environments to date. Our quantification of the refractory fraction of phosphorus in protoplanetary discs provides a reference point for planet formation and astrochemistry models, which require priors on what fraction (if any) of a given element is available for gas-phase chemistry. For P, we conclude this percentage is $\lesssim\,$4\,\%.

The exact reservoirs of P and the abundance of these reservoirs in a protoplanetary disc are still unknown. Elemental P in the diffuse ISM is predominantly or entirely present as an atomic gas, which depletes with increasing ISM density \citep{2009Jenkins}. Consistent with this trend, our results show that in the innermost regions of evolved protoplanetary discs, $\geq96\%$ of P is locked in solids. In line with the findings of \citet{2025Kama} for HD\,100546, these solids are relatively refractory, as they correlate strongly with dust-forming elements across a sample of systems with relatively warm dust traps \citep[e.g. $100$--$200\,$K for HD\,100546 and HD\,169242; see][]{2016Kama, 2024Keyte}. The P in dust has to be locked in minerals with a higher condensation temperature than water since the refractory fraction of O is zero within uncertainties \citep{2019Kama}, and most O in the protoplanetary disc is locked in water ice. \cite{2025Kama} found that the major P-carrying reservoir is consistent with apatite or schreibersite, with ammonium phosphates likely to be a less important carrier because of the lack of evidence for the depletion of N.

Future observations with higher S/N UV spectra will reduce uncertainties on P abundances and will provide an estimate of the refractory fraction of P. With lower uncertainties on P abundances, veiling and NLTE corrections will also become essential. Together with higher S/N UV spectra, increasing the sample size will also help improve constraints on the refractory fraction of phosphorus in the inner disc. 

\begin{acknowledgements}
This work has made use of the VALD database, operated at Uppsala University, the Institute of Astronomy RAS in Moscow, and the University of Vienna. This research is based on observations made with the NASA/ESA Hubble Space Telescope obtained from the Mikulski Archive for Space Telescopes (MAST) at the Space Telescope Science Institute, which is operated by the Association of Universities for Research in Astronomy, Inc., under NASA contract NAS 5–26555. These observations are associated with programs: 14703 (PI: Andrea Banzatti), 12996 (PI: Johns-Krull), 8300 (PI: Anne-Marie Lagrange). Based on observations collected at the European Southern Observatory under ESO programmes 077.D-0092(A), 082.D-0061(A), 084.A-9016(A), 085.A-9027(B). 
All authors gratefully acknowledge funding from the European Union's Horizon Europe research and innovation programme under grant agreement No. 101079231 (EXOHOST) and from UK Research and Innovation (UKRI) under the UK government’s Horizon Europe funding guarantee (grant number 10051045). AA acknowledges support from the Estonian Research Council grant PRG 2159. We thank the anonymous referee whose comments improved the quality of our paper.
\end{acknowledgements}

\bibliography{ref}      
\bibliographystyle{aa}

\begin{appendix}
\onecolumn
\section{Properties of the sample stars and their spectra}
\begin{table*}[h]
\renewcommand{\arraystretch}{1.2}
\tiny
\caption{Details about the optical and UV spectra of our sample stars.}
\begin{tabular}{l l l l l l l l l l}
\hline
\hline
Star & Optical & Optical spectral & Optical S/N & UV & UV spectral & UV S/N \\
 & spectrograph & resolution & (at $\sim5800\,\AA$ per spectral bin) & spectrograph & resolution & (at $\sim1700\,\AA$ per spectral bin) \\
\hline

HD\,36112 & ESPADONS & 80\,000 & 300 & COS & 20\,000 & 20\\

HD\,144432 & ESPADONS & 80\,000 & 373 & COS & 20\,000 & 10\\

HD\,169142 & ESPADONS & 80\,000 & 309 & STIS & 45\,800 & 10\\

HD\,139614 & ESPADONS & 80\,000 & 342 & STIS & 45\,800 & 10\\

HD\,100546 & FEROS & 48\,000 & 580 & STIS & 114\,000 & 20\\

\hline
\end{tabular}
\label{tab:sample_info_table}
\end{table*}

\begin{table*}[h]
\renewcommand{\arraystretch}{1.2}
\tiny
\caption{Stellar parameters, Fe and P abundance of our sample stars.}
\begin{tabular}{l l l l l l l l l}
\hline
\hline
Star & Type & Age$^1$\,(Myrs) & \teff\,(K) & \logg & \vsini\,(km\,s$^{-1}$) & \vmic\,(km\,s$^{-1}$) & $\log\,\rm (Fe/H)_*$ & $\log\,\rm (P/H)_*$ \\
\hline

HD\,36112 & HAeBe star & $10.92_{-0.97}^{+0.23}$ & $8000 \pm 100$ & $3.90 \pm 0.09$ & $57.7 \pm 0.3$ & $3.1 \pm 0.5$ & $-4.45 \pm 0.02$ & $-6.57_{-0.20}^{+0.20}$ \\

HD\,144432 & HAeBe star & $7.98_{-0.21}^{+0.02}$ & $7300 \pm 100$ & $3.52 \pm 0.11$ & $81.9 \pm 0.6$ & $3.4 \pm 0.1$ & $-4.64 \pm 0.04$ & $-6.74_{-0.22}^{0.13}$ \\

HD\,169142 & HAeBe star & $20$ & $7500 \pm 100$ & $4.09 \pm 0.17$ & $52.2 \pm 0.5$ & $2.4 \pm 0.6$ & $-5.07 \pm 0.05$ & $-7.41_{-1.0}^{0.23}$\\

HD\,139614 & HAeBe star & $19.35_{-0.0}^{+0.64}$ & $7600 \pm 100$ & $3.81 \pm 0.05$ & $25.6 \pm 0.2$ & $3.7 \pm 0.5$ &  $-5.02 \pm 0.04$ & $-7.28_{-0.15}^{0.11}$ \\

HD\,100546 & HAeBe star & $7.67_{-0.67}^{+0.36}$ & $10600 \pm 300$ & $4.17 \pm 0.35$ & $66.5 \pm 5.7$ & $2.0 \pm 0.1$ &  $-5.52 \pm 0.12$ & $-8.03_{-0.36}^{0.23}$ \\

HD\,37356$^2$ & YABS & $20.2 \pm 0.6$ & $22039 \pm 131$ & $3.96 \pm 0.02$ & $18.6 \pm 0.6$ & $1.6 \pm 0.8$ & $-4.93 \pm 0.06$ & $-7.06 \pm 0.07$\\
\hline
Sun$^3$ & & & & & & & $-4.54 \pm 0.04$ & $-6.59 \pm 0.03$ \\
\hline
\end{tabular}
\tablebib{1 - \cite{2021GuzmanDiaz}, 2 - \cite{2026Olbermann}, 3 - \cite{2021Asplund}}
\label{tab:sample_parameter_table}
\end{table*}

\begin{table*}[h]
\renewcommand{\arraystretch}{1.2}
\tiny
\caption{Chemical abundances of the sample stars.}
\begin{tabular}{l l l l l l l l l}
\hline
\hline
Star &  $\log\,\rm (C/H)_*$ & $\log\,\rm (N/H)_*$ & $\log\,\rm (O/H)_*$ & $\log\,\rm (Na/H)_*$ & $\log\,\rm (Mg/H)_*$ & $\log\,\rm (Al/H)_*$ & $\log\,\rm (Si/H)_*$ & $\log\,\rm (S/H)_*$ \\
\hline

HD\,36112 & $-3.51 \pm 0.04$ & $-4.20 \pm 0.20$ & $-3.13 \pm 0.11$ & $-5.61 \pm 0.20$ & $-4.33 \pm 0.20$ & $-5.37 \pm 0.20$ & $-4.26 \pm 0.10$ & $-4.82 \pm 0.03$\\

HD\,144432 & $-3.70	\pm 0.10$ & $-4.08 \pm 0.20$ & $-3.22 \pm 0.18$ & $-5.79 \pm 0.20$ & $-4.41 \pm 0.25$ & $-5.80 \pm 0.08$ & $-4.37 \pm 0.10$ & $-4.87 \pm 0.05$ \\


HD\,169142 & $-3.51 \pm 0.08$ & $-4.07 \pm 0.20$ & $-3.27 \pm 0.03$ & $-6.17 \pm 0.20$ & $-4.91 \pm 0.26$ & $-5.83 \pm 0.23$ & $-5.13 \pm 0.09$ & $-5.39 \pm 0.06$ \\

HD\,139614 & $-3.68 \pm 0.07$ & $-4.12 \pm 0.09$ & $-3.28 \pm 0.01$ & $-6.10 \pm 0.01$ & $-4.75 \pm 0.20$ & $-5.80 \pm 0.22$ & $-4.82 \pm	0.14$ & $-5.28 \pm 0.08$ \\

HD\,100546 & $-3.49 \pm 0.20$ & $-3.75 \pm 0.14$ & $-3.24 \pm 0.11$ & & $-5.37 \pm 0.25$ & $-6.56 \pm 0.20$ & $-5.22 \pm 0.25$ \\
\hline
Sun$^1$ & $-3.54 \pm 0.04$ & $-4.17 \pm 0.07$ & $-3.31 \pm 0.04$ & $-5.78 \pm 0.03$ & $-4.45 \pm 0.03$ & $-5.57 \pm 0.03$ & $-4.49 \pm 0.03$ & $-4.88 \pm 0.03$ \\
\hline \\ 

\hline
\hline
Star &  $\log\,\rm (Ca/H)_*$ & $\log\,\rm (Sc/H)_*$ & $\log\,\rm (Ti/H)_*$ & $\log\,\rm (V/H)_*$ & $\log\,\rm (Cr/H)_*$ & $\log\,\rm (Mn/H)_*$ & $\log\,\rm (Ni/H)_*$ & $\log\,\rm (Cu/H)_*$ \\
\hline

HD\,36112 & $-5.49 \pm 0.06$ & $-8.85 \pm 0.03$ & $-6.97 \pm 0.10$ & $-7.98 \pm 0.20$ & $-6.24	\pm 0.06$ & $-6.36 \pm 0.09$ & $-5.71 \pm 0.25$ & $-7.93 \pm 0.20$ \\

HD\,144432 & $-5.75 \pm 0.13$ & $-9.18 \pm 0.14$ & $-7.26 \pm 0.10$ & $-8.16 \pm 0.20$ & $-6.51 \pm 0.11$ & $-6.69 \pm 0.16$ & $-6.00 \pm 0.15$ & $-8.05 \pm 0.20$\\


HD\,169142 & $-6.21 \pm 0.04$ & $-9.50 \pm 0.05$ & $-7.62 \pm 0.10$ & $-8.86 \pm 0.20$ & $-6.97 \pm 0.07$ & $-7.01 \pm 0.16$ & $-6.49 \pm 0.02$ & $-8.44 \pm 0.20$ \\

HD\,139614 & $-6.11 \pm 0.03$& $-9.40 \pm 0.02$ & $-7.51 \pm 0.05$ & $-8.60 \pm 0.20$ & $-6.82 \pm 0.07$ & $-7.16 \pm 0.09$ & $-6.34 \pm 0.07$ & $-8.35 \pm 0.20$ \\

HD\,100546 & $-6.47 \pm 0.20$ & & $-8.41 \pm 0.13$ & $-7.44 \pm 0.05$ & &  \\
\hline
Sun$^1$ & $-5.70 \pm 0.03$ & $-8.86 \pm 0.04$ & $-7.03 \pm 0.05$ & $-8.10 \pm 0.08$ & $-6.38 \pm 0.04$ & $-6.58 \pm 0.06$ & $-5.80 \pm 0.04$ & $-7.82 \pm 0.05$ \\
\hline \\ 

\hline
\hline
Star &  $\log\,\rm (Zn/H)_*$ & $\log\,\rm (Y/H)_*$ & $\log\,\rm (Ba/H)_*$ & $\log\,\rm (La/H)_*$ & $\log\,\rm (Nd/H)_*$\\
\hline

HD\,36112 & $-7.84 \pm 0.20$ & $-9.76 \pm 0.20$ & $-9.45 \pm 0.30$ & $-10.74 \pm 0.20$ \\

HD\,144432 & $-8.03 \pm 0.20$ & $-9.93 \pm 0.20$ & $-9.68 \pm 0.22$ & $-11.56 \pm 0.20$ \\

HD\,169142 & $-8.14 \pm 0.20$ & $-10.39 \pm 0.20$ & $-10.12 \pm 0.12$ &  & $-10.64 \pm 0.49$\\

HD\,139614 & $-8.13 \pm 0.20$ & $-10.29 \pm 0.20$ & $-10.19 \pm	0.10$ & $-11.58 \pm 0.20$ & $-10.66 \pm 0.18$ \\

HD\,100546 & & & $-10.20 \pm 0.20$ &  \\
\hline
Sun$^1$ & $-7.44 \pm 0.05$ & $-9.79 \pm 0.05$ & $-9.73 \pm 0.05$  & $-10.89 \pm 0.04$ & $-10.58 \pm 0.04$ \\
\hline
\end{tabular}
\tablebib{1 - \cite{2021Asplund}.}
\label{tab:sample_abundance_table}
\end{table*}

\twocolumn

\section{Optical spectral analysis}\label{appendix:optical analysis}
\begin{figure}[h]
\resizebox{\hsize}{!}{\includegraphics[trim=0.2cm 0.0cm 0.0cm 0.2cm]{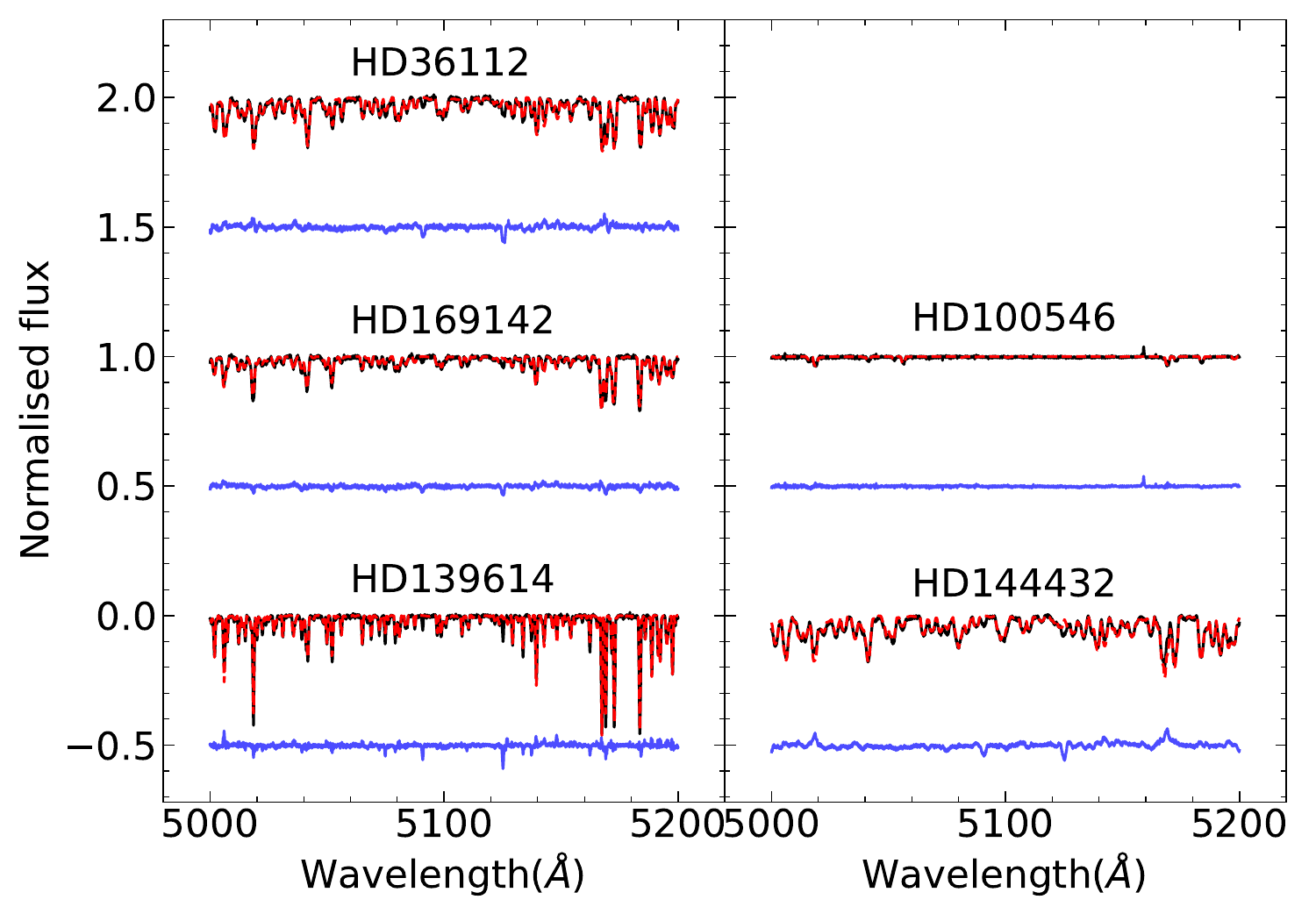}}
 \caption{Comparison between synthetic spectra (red) generated using the estimated stellar parameters and abundances and observed spectra (black) and difference between the two spectra (blue). The spectra for each star are shifted vertically for better visualisation.}
 \label{fig:optical_PPD_fit}
\end{figure}
We used the {\sc zeeman} spectral synthesis code \citep{1988Landstreet,2001Wadeetal,2012Folsom} together with linelist from the Vienna Atomic Line Database\footnote{\texttt{\href{https://vald.astro.uu.se/}{https://vald.astro.uu.se/}}} \citep[VALD;][]{1995Piskunovetal,1997Ryabchikovaetal,1999Kupka,2000Kupkaetal,2015Ryabchikova} and {\sc atlas9} model atmospheres \citep{1993Kurucz,2004castelliKurucz} for spectral synthesis and fitting of optical and UV spectra. We followed the spectral fitting method described also in \cite{2025Borthakur} to estimate stellar parameters and abundances. The spectral fitting method works by comparing a synthetic spectrum generated from a set of parameters with the observed spectrum and varying the parameters until the $\chi^2$ is minimised.

The optical spectra of the sample stars were continuum normalised using {\sc suppnet} \citep{2022suppnet}. We estimated the following stellar parameters from the optical spectra: effective temperature (\teff), surface gravity (\logg), projected rotational velocity (\vsini), microturbulence velocity (\vmic) and radial velocity (\vr).  All HAeBe stars in our sample have literature values, which we used as the initial stellar parameters and abundances. We used the stellar parameters and abundances from \cite{2012Folsom} for all HAeBe stars, except HD\,100546 which is taken from \cite{2016Kama}.  

For the metal line analysis, we used the optical spectral window between 4400\,$\AA$ and 7000\,$\AA$, excluding regions with emission lines, tellurics, and Balmer lines. Using the initial stellar parameters and solar abundances from the previous step, we fit the metal lines for all stellar parameters and elemental abundances of the dominant lines (e.g., Fe, Ti, Cr). We used these stellar parameters and abundance estimates for the next step, to also obtain uncertainties. We divided the selected spectral window into four to five approximately equal-width spectral windows and fit each window separately. We simultaneously fit all stellar parameters and elemental abundances of the dominant lines in each wavelength window. We reported the average value across these windows as the stellar atmospheric parameters and abundances, and the standard deviation of the scatter between the wavelength windows is reported as their uncertainties. For HD\,100546, spectral fitting failed to estimate \vmic. We calculated its \vmic\ value using the formula by \cite{2006Pace} 
\begin{equation}
    \nu_{\rm mic} = -4.7\log(T_{\rm eff}) + 20.9\, \rm km\,s^{-1}.
    \label{eqn:vmic_equation}
\end{equation}
Finally, we estimated the abundances of other elements from individual spectral lines. Figure\,\ref{fig:optical_PPD_fit} shows the comparison of observed spectra of our sample stars with their synthetic spectra calculated with the estimated parameters. Table\,\ref{tab:sample_parameter_table} and Table\,\ref{tab:sample_abundance_table} list the stellar parameter and abundances for our sample stars and Table\,\ref{tab:wavelength regions for elements} lists the wavelength regions used for each element. If the abundance was measured from multiple lines, then the average value is reported, with the scatter in their abundances as the uncertainty. If the abundance is measured from a single line, we assigned a typical uncertainty of 0.20\,dex. 

The stellar parameters of all the stars analysed in this paper match with literature values within one standard deviation except for \logg. The average difference between our \logg\ estimates and the literature values is $-0.17\,\pm\,0.14$.

\section{UV spectral analysis}\label{appendix:uv analysis}
\begin{figure*}
\includegraphics[width=\linewidth]{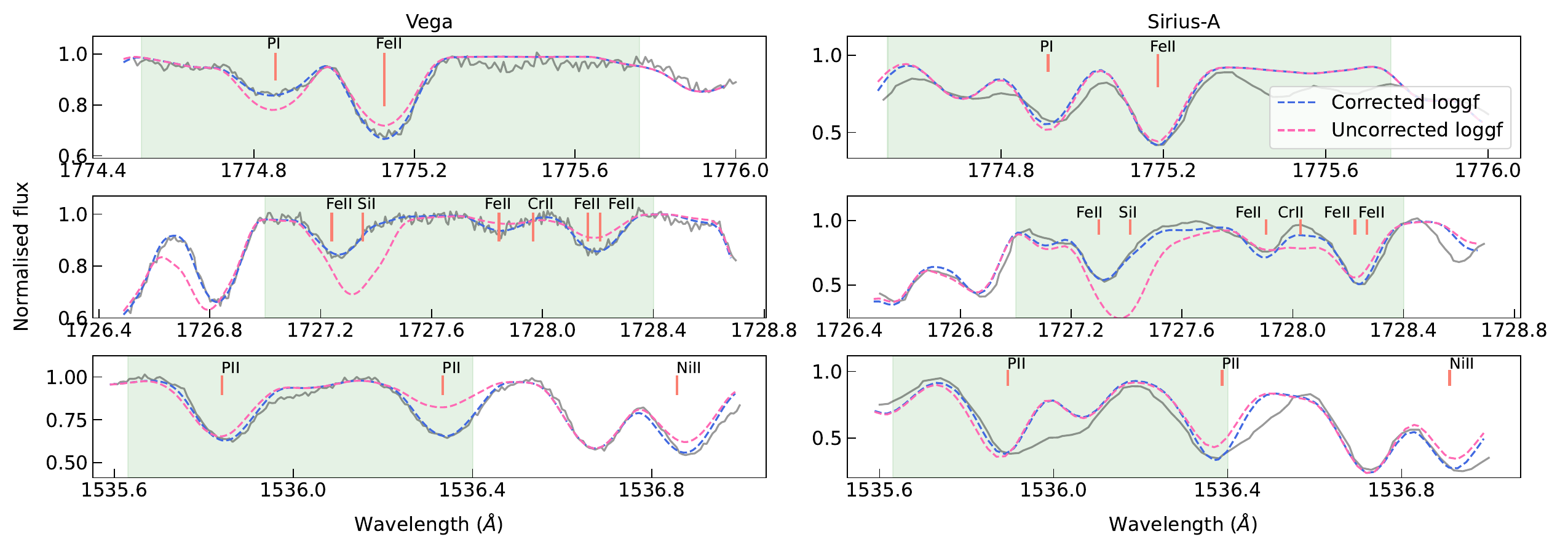}
 \caption{Comparing observed spectrum of Vega (left column) and Sirius-A (right column) with synthetic spectrum generated with literature and corrected \loggf\ values. The green-highlighted region is the wavelength range used to estimate the P-abundances for the sample stars.}
 \label{fig:Vega_Sirirus_loggf}
\end{figure*}

We corrected the oscillator strengths, \loggf\ of some of the lines around the P-lines of interest using the Sirius and Vega spectra with their stellar parameter and abundances from \cite{2011Landstreet} and \cite{2010Vega}, respectively. Table\,\ref{tab:P-line loggf corrections} lists the corrected \loggf\ value and the value reported in VALD and Figure\,\ref{fig:Vega_Sirirus_loggf} show the comparison before and after \loggf\ corrections for both Vega and Sirius-A.

\begin{figure}
\includegraphics[width=\linewidth]{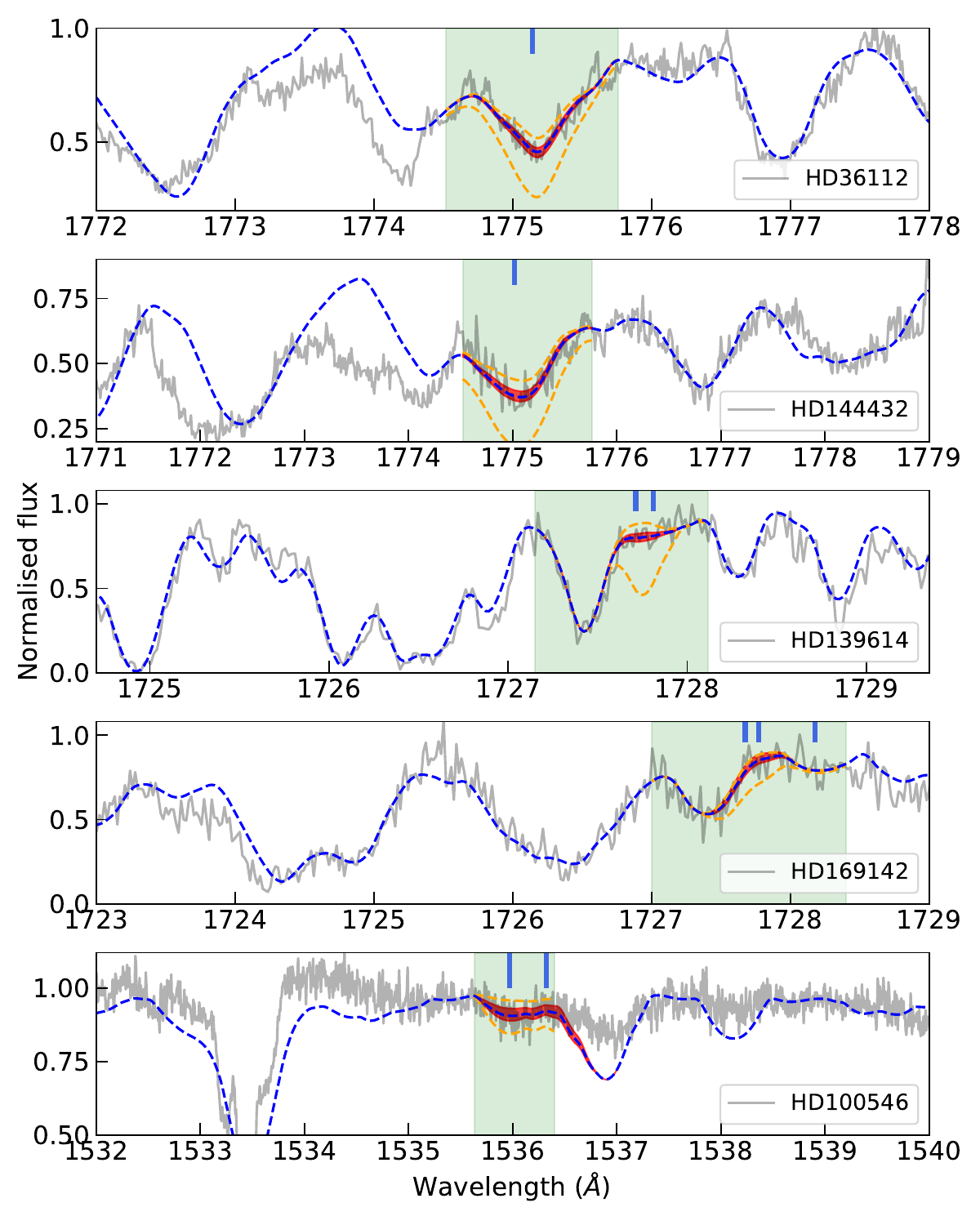}
 \caption{UV spectral fits of the sample stars for the P abundance estimate. The light blue vertical lines represent the P-lines in that spectral region. The blue dashed line is the best fit, and the red region is the spectrum with uncertainty on the P abundance from the 5th and 95th percentiles of the posterior distribution. The green highlighted region is fit to estimate the P-abundance. The orange spectra represent 1 dex lower and higher P abundance. The P abundances and the uncertainties for the sample stars are listed in Table\,\ref{tab:sample_P_abundance_table}.}
 \label{fig:P_abundance_fit_plot}
\end{figure}

\begin{figure}[h]
    \centering
    \includegraphics[width=0.8\linewidth]{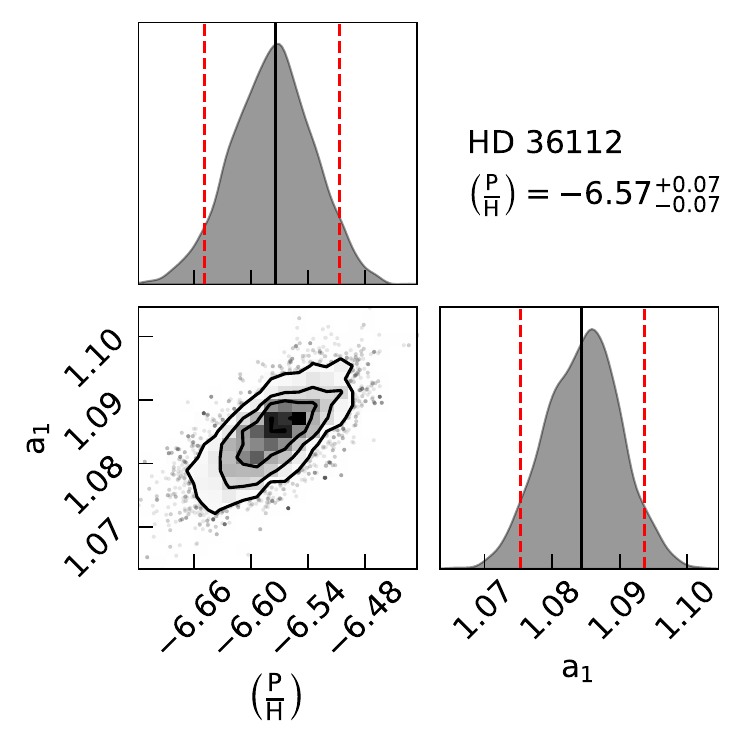}
    \caption{MCMC posterior distribution for HD\,36112 P abundance fitting. The black vertical lines represent the peak of the posterior and the red lines represent 5th and 95th percentile of the posteriors.}
    \label{fig:HD36112_pdf}
\end{figure}

\begin{figure}[h]
    \centering
    \includegraphics[width=0.8\linewidth]{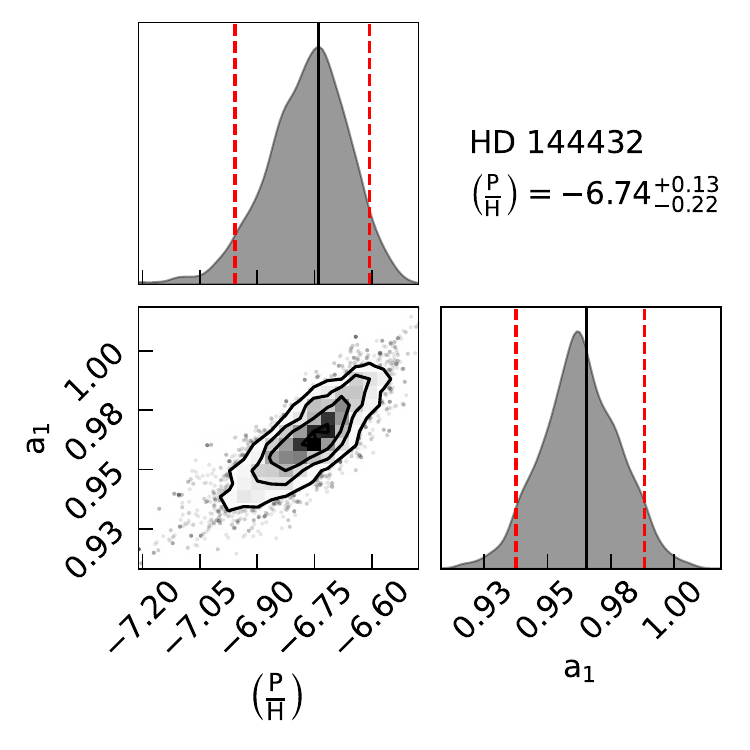}
    \caption{Same as Figure\,\ref{fig:HD36112_pdf} for HD\,144432.}
    \label{fig:HD144432_pdf}
\end{figure}

\begin{figure}[h]
    \centering
    \includegraphics[width=0.8\linewidth]{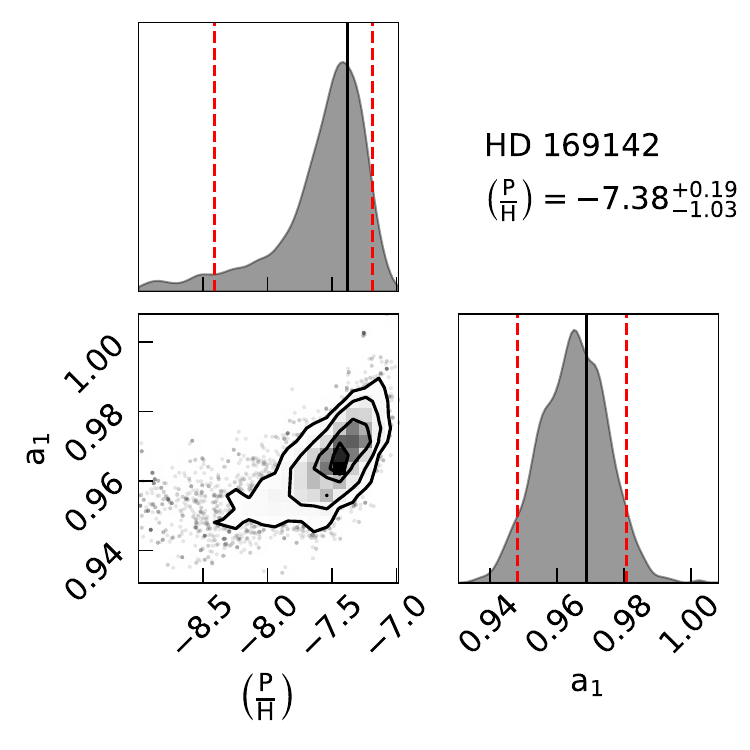}
    \caption{Same as Figure\,\ref{fig:HD36112_pdf} for HD\,169142. }
    \label{fig:HD169142_pdf}
\end{figure}

\begin{figure}[h]
    \centering
    \includegraphics[width=0.8\linewidth]{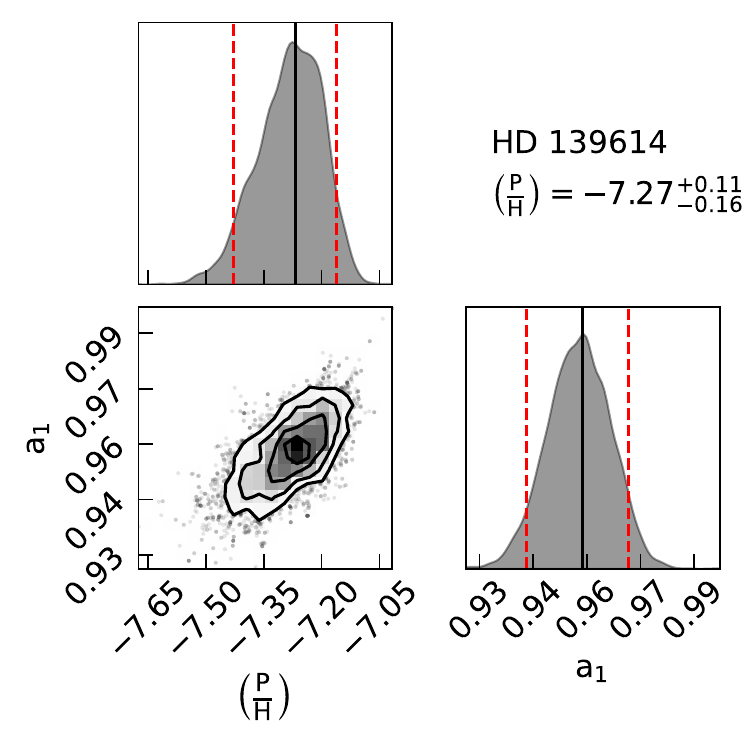}
    \caption{Same as Figure\,\ref{fig:HD36112_pdf} for HD\,139614.}
    \label{fig:HD139614_pdf}
\end{figure}

\begin{figure}[h]
    \centering
    \includegraphics[width=0.8\linewidth]{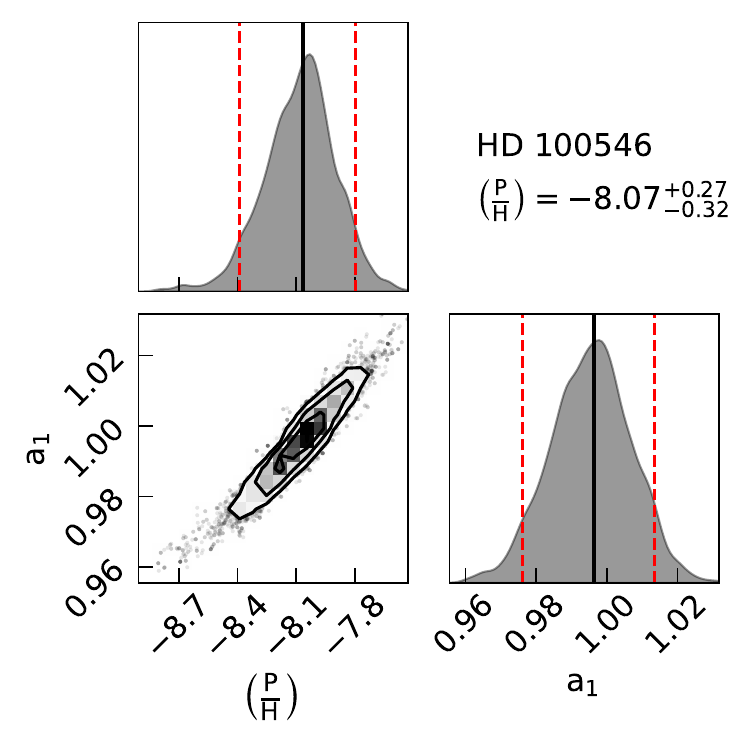}
    \caption{Same as Figure\,\ref{fig:HD36112_pdf} for HD\,100546.}
    \label{fig:Hd100546_pdf}
\end{figure}

For the P abundance analysis, we first normalised the UV spectrum using \textsc{suppnet} by comparing with a synthetic spectrum generated using the estimated stellar parameters, stellar abundances and the \textsc{zeeman} spectral synthesis code. We consider small spectral regions around the UV P lines and simultaneously fit for a constant scaling factor (zeroth-degree polynomial) and the P abundance using an MCMC fitting routine. We used a uninformed prior for the scaling factor in $[-\infty,+\infty]$ and a uniform prior for $\rm \log\left(\frac{P}{H}\right)$ in $[-8,-6]$. We reported the 5th and 95th percentiles of the posterior distribution of $\rm \log\left(\frac{P}{H}\right)$ as the uncertainties and the peak of the posterior as the actual $\rm \log\left(\frac{P}{H}\right)$ for the star. The P-abundance is listed in Table\,\ref{tab:sample_parameter_table}. 
The exact P-lines used for all the sample stars are listed in Table\,\ref{tab:sample_P_abundance_table}. We used P~{\sc i} lines for all stars except HD\,100546, for which we used a blended region containing two P~{\sc ii} lines.  For HD\,36112, the estimated uncertainty on the P-abundance was $\pm 0.07$ dex, which seemed to be underestimated, and thus, we assign the uncertainty on the inferred P-abundance to be $\pm\,0.20$ dex. The UV spectral fits are shown in Figure\,\ref{fig:P_abundance_fit_plot}. The posterior distribution of the P-abundance fit for all sample stars is shown in Figure\,\ref{fig:HD36112_pdf}, \ref{fig:HD144432_pdf}, \ref{fig:HD169142_pdf},\ref{fig:HD139614_pdf}, and \ref{fig:Hd100546_pdf}.

\begin{table}[h]
\renewcommand{\arraystretch}{1.2}
\tiny
\caption{Line corrections.}
\begin{tabular}{l l l l l}
\hline
\hline
Line & Wavelength ($\AA$) & \loggf$_{\mathrm{old}}$ & \loggf$_{\mathrm{new}}$ & Ref. \\
\hline
P~\textsc{ii} & 1535.923 & -1.765 & -1.470 & \cite{K12} \\
P~\textsc{ii} & 1536.416 & -2.274 & -1.600 & \cite{K12} \\
Ni~\textsc{ii} & 1536.939 & -1.631 & -1.000 & \cite{K03} \\
Fe~\textsc{ii} & 1727.3329 & -0.121 & -0.244 & \cite{K13} \\
Si~\textsc{i}  & 1727.4454 & -1.057 & -2.013 & \cite{K07} \\
Fe~\textsc{ii} & 1727.9360 & -1.575 & -1.300 & \cite{K13} \\
Cr~\textsc{ii} & 1728.0600 & -0.605 & -10.349 & \cite{K16} \\
Fe~\textsc{ii} & 1728.2569 & -2.463 & -1.852 & \cite{K13} \\
Fe~\textsc{ii} & 1728.3005 & -3.006 & -2.734 & \cite{K13} \\
P~\textsc{i} & 1774.949 & -0.210 & -0.460 & \cite{LAW} \\
Fe~\textsc{ii} & 1775.220 & -2.943 & -2.535 & \cite{K13} \\
\hline
\end{tabular}
\tablefoot{All the line data are taken from the VALD database.}
\label{tab:P-line loggf corrections}
\end{table}

\begin{table}[h]
    \renewcommand{\arraystretch}{1.2}
    \small
    \caption{Lines used to estimate the P-abundance for the sample stars.}
    \begin{tabular}{l l l}
       \hline
       \hline
       Star & Wavelength region & P-lines ($\AA$) \\
       \hline
       HD\,36112 & $1774.52-1775.76$ & $1774.949$\\       
       
       HD\,144432 & $1774.52-1775.76$ & $1774.949$\\

        HD\,169142 & $1727-1728.4$ & $1727.716, 1727.815$ \\
       
       HD\,139614 & $1727.15-1728.11$ & $1727.716, 1727.815$ \\
       
       HD\,100546 & $1535.63-1536.4$ & $1535.923^*, 1536.416^*$\\
        \hline
    \end{tabular}
    \tablefoot{All the P-lines listed in this table are P\,{\sc i} lines except the lines marked with $^*$, which are P\,{\sc ii} lines.}
    \label{tab:sample_P_abundance_table}
\end{table}

\section{Posterior distribution for refractory fraction of P}
\begin{figure}[h]
    \centering
    \includegraphics[width=\linewidth]{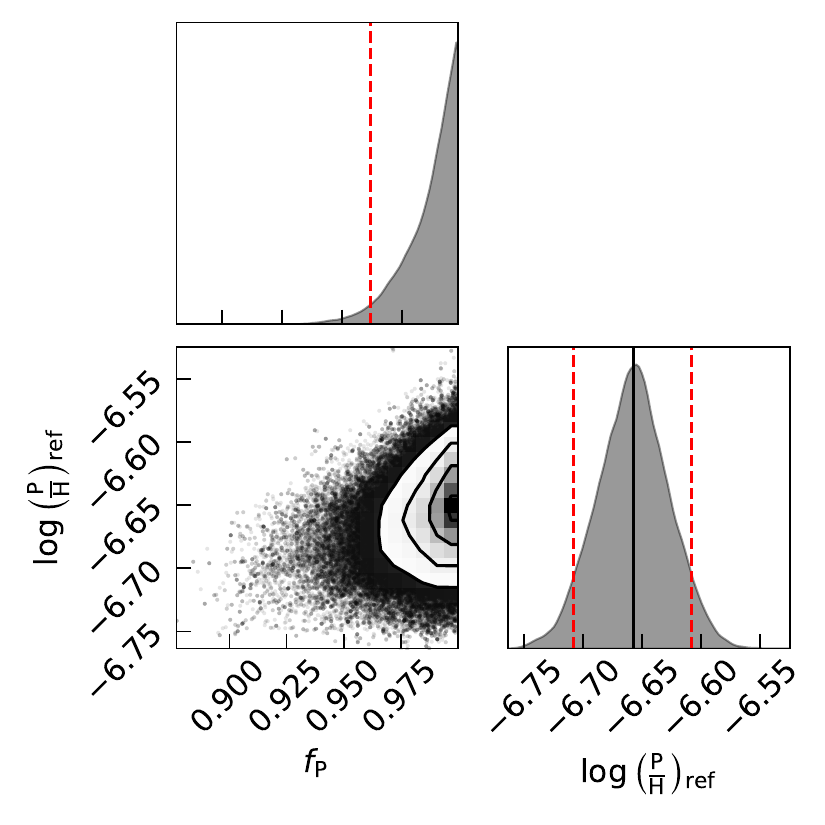}
    \caption{Corner plot of the posterior distribution estimated from the MCMC fit to estimate the refractory fraction of phosphorus and the reference phosphorus abundance. The red vertical line at the top plot represents the 5th percentile of the posterior distribution of $f_{\rm P}$. The two red vertical lines in the bottom right plot represent the 5th and 95th percentiles of the posterior distribution of $\log\rm \left(\frac{P}{H}\right)_{ref}$, and the black vertical line represents the peak of the distribution.}
    \label{fig:corner_plot_pfrac_pref}
\end{figure}

\section{List of all spectral lines}
\begin{table}[h]
    \renewcommand{\arraystretch}{1.5}
    \tiny
    \caption{Wavelength regions used to estimate elemental abundances of the sample stars.}
    \begin{tabular}{l l}
    \hline
    \hline
    Element & Wavelengths ($\AA$) \\
    \hline
    C &  $(5009.58-5046.12),	(5044.86-5061.03), $ \\
      &  $(5376.29-5419.75), (6818.52-6833.6), (7082.86-7127.28)$\\
    \hline
    N &  $(7457.49-7488.37)$\\
    \hline
    O &  $(4961.62-4976.4),	(5320.21-5346.99),$ \\
    & $ (5419.7-5441.6), (6143.19-6175.35),	(6447.74-6462.61)$\\
    \hline
    Na &  $(5676.17-5696.21)$\\
    \hline
    Mg &  $(4421.65-4439.53), (4458.5-4498.29),	(4689.22-4719.38),$ \\
    & $ (4725.37-4748.33), (5157.58-5177.01), (5176.62-5201.83), $\\
    & $(5244.61-5280.33), (5387.69-5419.93), (5518.89-5540.75)$\\
    \hline
    Al & $(4659.74-4676.65)$ \\
    \hline
    Si &  $(5044.89-5062.09), (6363.46-6377.9)$\\
    \hline
    S &  $(6738.57-6752.93), (6752.93-6762.05)$\\
    \hline
    Ca & $(4420.79-4439.2),	(4447.08-4465.49),	(4568.7-4607.05), $\\ 
    & $ (4997.02-5029.65), (5249.85-5280.66),	(5579.88-5609.06)$ \\
    \hline
    Sc & $(4659.74-4676.65), (5024.52-5045.7),$\\
    & $ (5517.77-5539.75), (5651.53-5676.04)$ \\
    \hline
    Ti &  $(4407.42-4439.04), (4438.25-4498.48), (4511.53-4538.54),$\\
    & $ (4538.07-4569.22), (4568.89-4608.21),	(4650.88-4661.17),$\\
    & $(4996.74-5045.95), (5060.15-5105.09),	(5114.68-5143.45),$\\
    & $(5177.09-5221.98), (5320.48-5354.68)$\\
    \hline
    V &  $(4517.55-4537.96)$\\
    \hline
    Cr &  $(4537.27-4569.27), (4607.87-4641.52), (5221.37-5289.2),$\\
    & $ (5288.45-5320.81),	(5320.42-5355.04),$\\
    & $ (5470.66-5484.77), (5497.37-5517.25)$\\
    \hline
    Mn & $(4447.16-4475.15), (4751.41-4759.25),	(4759.25-4776.48)$ \\
    \hline
    Ni &  $(4387.93-4408.87), (4699.0-4721.43),	(5026.19-5045.74)$\\
    \hline
    Cu & $(5102.49-5113.89)$ \\
    \hline
    Zn & $(4684.59-4725.35)$ \\
    \hline
    Y & $(5084.2-5091.33)$ \\
    \hline
    Ba & $(4544.65-4568.94), (6130.91-6153.18)$ \\
    \hline
    La &  $(4517.55-4537.96)$\\
    \hline
    Nd & $(4452.58-4467.83), (5245.27-5252.6)$ \\
    \hline
    \end{tabular}
    \tablefoot{The choice of the wavelength regions depends on the wavelength coverage of the optical spectrum for each sample star.}
    \label{tab:wavelength regions for elements}
\end{table}
\end{appendix}
\end{document}